\documentclass[acmtog]{acmart}
\acmSubmissionID{1234}

\usepackage{booktabs} % For formal tables

\usepackage[ruled]{algorithm2e} % For algorithms

\SetAlFnt{\small}
\SetAlCapFnt{\small}
\SetAlCapNameFnt{\small}
\SetAlCapHSkip{0pt}

\def\rest{\textbf{X}}
\def\deformed{\textbf{x}}
\def\energy{\mathcal{E}}
\def\kernel{\Psi}
\def\volume{A}
\def\stretch{\sigma}
\def\dg{\textbf{F}}
\def\hessian{\textbf{H}}
\def\projection{\textbf{K}}
\def\eigen{\textbf{D}}
\def\feasible{\mathcal{C}}
\def\search{\textbf{p}}
\def\diagonal{\boldsymbol{\Lambda}}
\def\systemhessian{\boldsymbol{\mathcal{H}}}

\makeatletter
\newcommand*{\rom}[1]{\expandafter\@slowromancap\romannumeral #1@}
\makeatother

\begin{document}
% Title portion
\title{Eigen Space of Mesh Distortion Energy Hessian}

% DO NOT ENTER AUTHOR INFORMATION FOR ANONYMOUS TECHNICAL PAPER SUBMISSIONS TO SIGGRAPH 2019!
\author{Yufeng Zhu}
%\orcid{1234-5678-9012-3456}
%\affiliation{%
  %\institution{Ziva Dynamics}
  %\streetaddress{319 W Pender St}
  %\city{Vancouver}
  %\state{BC}
  %\postcode{V6B 1T3}
  %\country{Canada}}
\email{mike323zyf@gmail.com}
%\affiliation{%
%  \institution{Inria Paris-Rocquencourt}
%\email{beranger@inria.fr}
%\affiliation{%
% \institution{Rajiv Gandhi University}
%\email{aprna_patel@rguhs.ac.in}
%\affiliation{%
%  \institution{Tsinghua University}
%  \streetaddress{30 Shuangqing Rd}
%  \country{China}
%}
%\email{chan0345@tsinghua.edu.cn}
%\author{Ting Yan}
%\affiliation{%
%  \institution{Eaton Innovation Center}
%  \city{Prague}
%  \country{Czech Republic}}
%\email{yanting02@gmail.com}
%\author{Tian He}
%\affiliation{%
%  \institution{University of Virginia}
%  \department{School of Engineering}
%  \city{Charlottesville}
%  \state{VA}
%  \postcode{22903}
%  \country{USA}
%}
%\affiliation{%
%  \institution{University of Minnesota}
%\email{tinghe@uva.edu}
%\author{Chengdu Huang}
%\author{John A. Stankovic}
%\author{Tarek F. Abdelzaher}
%\affiliation{%
%  \institution{University of Virginia}
%  \department{School of Engineering}
%  \city{Charlottesville}
%  \state{VA}
%  \postcode{22903}
%  \country{USA}
%}

%\renewcommand\shortauthors{Zhou, G. et al}

\begin{abstract}
Mesh distortion optimization is a popular research topic and has wide range of applications in computer graphics, including geometry modeling, variational shape interpolation, UV parameterization, elastoplastic simulation, etc. In recent years, many solvers have been proposed to solve this nonlinear optimization efficiently, among which projected Newton has been shown to have best convergence rate and work well in both 2D and 3D applications. Traditional Newton approach suffers from ill conditioning and indefiness of local energy approximation. A crucial step in projected Newton is to fix this issue by projecting energy Hessian onto symmetric positive definite (SPD) cone so as to guarantee the search direction always pointing to decrease the energy locally. Such step relies on time consuming Eigen decomposition of element Hessian, which has been addressed by several work before on how to obtain a conjugacy that is as diagonal as possible. In this report, we demonstrate an analytic form of Hessian eigen system for distortion energy defined using principal stretches, which is the most general representation. Compared with existing projected Newton diagonalization approaches, our formulation is more general as it doesn't require the energy to be representable by tensor invariants. In this report, we will only show the derivation for 3D and the extension to 2D case is straightforward.
\end{abstract}

%\begin{CCSXML}
%<ccs2012>
%  <concept_id>10010520.10010553.10010562</concept_id>
%  <concept_desc>Computer systems organization~Embedded systems</concept_desc>
%  <concept_significance>500</concept_significance>
%  <concept_id>10010520.10010575.10010755</concept_id>
%  <concept_desc>Computer systems organization~Redundancy</concept_desc>
%  <concept_significance>300</concept_significance>
%  <concept_id>10010520.10010553.10010554</concept_id>
%  <concept_desc>Computer systems organization~Robotics</concept_desc>
%  <concept_significance>100</concept_significance>
%  <concept_id>10003033.10003083.10003095</concept_id>
%  <concept_desc>Networks~Network reliability</concept_desc>
%  <concept_significance>100</concept_significance>
%</ccs2012>
%\end{CCSXML}

%\ccsdesc[500]{Computer systems organization~Embedded systems}


%\keywords{Wireless sensor networks, media access control,
%multi-channel, radio interference, time synchronization}

\maketitle

\section{Introduction}

There have already been many interesting mesh distortion optimization solvers so far \cite{Teran:2005:RQF,Stomakhin:2012:ECI,Sin:2011:IIH,Xu:2015:NMD,Bouaziz:2014:PDF,Kovalsky:2016:AQP,Rabinovich:2017:SLI,Liu:2017:QMR,Shtengel:2017:GOV,Claici:2017:IPM,Zhu:2018:BCQ,Liu:2018:PP,Peng:2018:AAG,Smith:2018:SNF,Smith:2018:AEF}. As this is just a technical report on hessian diagonalization, we will only include the references here instead of going to detailed discussion one by one. The most related previous work are \cite{Teran:2005:RQF,Stomakhin:2012:ECI,Smith:2018:SNF,Smith:2018:AEF}. However, they either assume that the distortion energy can be represented using tensor invariants or just achieve a block diagonal structure. In this work, we will show the analytic eigen system of element hessian whose energy is defined using principal stretches, which is more general representation of mesh distortion. For any 3D mesh distortion energy defined using principal stretches, we prove that its diagonal form is composed of six scalars and one 3$\times$3 block. Depending on the energy definition, the 3$\times$3 block might also have an analytic diagonalization.

\section{Related Work}

\subsection{Distortion Energies}

Mesh distortion optimization is largely characterized by distortion energy. Many fundamental physical and geometric modeling problems reduce to minimizing measures of distortion over meshes. A wide range of energies have been proposed to fullfil such tasks for various application purposes. Linear approaches have the benefit of good performance as the distortion energies adopted are usually modeled as quadratic and require only a single linear system factorization~\cite{Lipman:2005:LRC,Weber:2009:CBC,Weber:2007:CAS,Zayer:2005:HGF}, but they also suffer from severe artifacts under large deformations. On the other hand, nonlinear measurement requires more sophisticated solvers but behaves well even for extreme deformations. 

Diverse range of nonlinear energies have been proposed to minimize various mapping distortions in the field of geometry processing, generally focused on minimizing either measures of isometric~\cite{Aigerman:2015:SSM,Chao:2010:SGM,Liu:2008:LAM,Smith:2015:BPF} or conformal~\cite{Ben:2008:CFB,Desbrun:2002:IPO,Hormann:2002:MIP,Levy:2002:LSC,Mullen:2008:SCP,Weber:2012:CEQ} distortion. As rigid as possible (ARAP), as similar as possible (ASAP) and as killing as possible (AKAP) are three alterative ways to minimize isometric/conformal distortions~\cite{Alexa:2000:ASI,Igarashi:2005:ASM,Sorkine:2007:ASM,Solomon:2011:AKA}. Other type energies like Dirichlet energy~\cite{Schueller:LIM:2013}, maximal stretch energy~\cite{Sorkine:2002:BDP} and Green-Lagrange energy~\cite{Bonet:2008:NCM} are also popular choices for applications. 

Distortion optimization is also applicable to physical based animation which typically minimizes hyperelastic potentials formed by integrating strain energy densities over the material domain to simulate elastic solids with large deformations. These material models date back to Mooney~\cite{Mooney:1940:ATO} and Rivlin~\cite{Rivlin:1948:SAO}. Their Mooney-Rivlin and Neo-Hookean materials, and many subsequent hyperelastic materials, e.g. St. Venant-Kirchoff, Ogden, Fung~\cite{Bonet:1998:ASO}, are constructed from empirical observation and analysis of deforming real-world materials. Material properties in these models are specified according to experiment for scientific computing applications~\cite{Ogden:1972:LDI}, or alternately are directly set by users in other cases~\cite{Xu:2015:NMD}, e.g., to meet artistic needs. Modified energy model~\cite{Stomakhin:2012:ECI} has also been proposed in computer graphics to stablize physical simulation.

There are various ways to parameterize the aforementioned distortion energies, including strain tensor invariants~\cite{Teran:2005:RQF}, stretch tensor invariants~\cite{Smith:2018:AEF}, principal stretches (singular values of mapping Jacobian)~\cite{Stomakhin:2012:ECI,Xu:2015:NMD}, etc. Among these options, principal stretch is the most general parameterization approach as it not only covers both 2D and 3D cases but also plays a more fundamental role than the others in defining distortion measurement. Moreover, it has been shown that principal stretch based distortion energy is more user-friendly and can be manipulated or modified for application purposes more easily and intuitively~\cite{Stomakhin:2012:ECI,Xu:2015:NMD}. 

\subsection{First Order Methods}

To obtain deformed mesh with optimal distortion measured by energies we just introduced, many solvers have been proposed and studied so far. The local-global method has been recognized as one of the most popular approaches and applied to many applications, including surface modeling~\cite{Sorkine:2007:ASM}, parameterization~\cite{Liu:2008:LAM} and volumetric deformation~\cite{Bouaziz:2014:PDF}, etc. Not until recent years did researchers find out that such method is closely related to Sobolev gradient~\cite{Neuberger:2006:SDF,Neuberger:2010:SGA}. Kovalsky et al.~\shortcite{Kovalsky:2016:AQP} extended this method by introducing acceleration to speed up its convergence, while Liu et al.~\shortcite{Liu:2017:QMR} instead combined it with L-BFGS for the same goal. Other notable solver improvement includes iterative reweighting Laplacian method~\cite{Rabinovich:2017:SLI} and killing vector field proximal algorithm~\cite{Claici:2017:IPM}. Both methods improve solver convergence by using more effective quadratic approximation during each iteration but also require solving a new linear system at every step. To overcome such issue, alternative approaches, which focus on efficient way of constructing effective local approximation, have been proposed by Zhu et al.~\shortcite{Zhu:2018:BCQ} and Peng et al.~\shortcite{Peng:2018:AAG}, both of which are variants of Broyden class method. All the listed techniques are regarded as first order methods as they rely on linear approximation of distortion energy. Other first order methods, like Gauss Newton~\cite{Eigensatz:2009:PMA}, block coordinate descent~\cite{Fu:2015:CLI} and L-BFGS~\cite{Smith:2015:BPF}, are also used in graphics applications, but they all share the poor convergence rate problem, which leads to the extensive study of efficient second order solvers.

\subsection{Second Order Methods}

Second order methods generally can achieve the most rapid convergence for convex energies but requires modification for nonconvex ones~\cite{Nocedal:2006:Book} to ensure that the proxy is at least positive semi-definite (PSD). At each iterate, the distortion energy hessian is evaluated to form a proxy matrix and special care must be taken in order to handle its indefiness. Trust region could be a quick fix to this problem and actually has been adopted by several previous work~\cite{Chao:2010:SGM,Schueller:LIM:2013}. However, finding the proper window size could require much more extra computational cost without any simplification, like the Dogleg method. Another popular strategy is fixed Newton which attracts lots of attention from research community in recent years. Composite majorization, a tight convex majorizer, was recently proposed as an analytic PSD approximation of the
hessian~\cite{Shtengel:2017:GOV}. Its proxy is efficient to assemble, but is limited to 2D problems. More general is the projected Newton method that projects per-element hessians to the PSD cone prior to assembly~\cite{Teran:2005:RQF,Stomakhin:2012:ECI,Xu:2015:NMD,Fu:2016:CIM}. However, these approaches still depend on numerical constructions of projected hessian and do not yield closed-form expressions for the underlying hessian's eigen pairs. Chen and Weber~\shortcite{Chen:2017:GLI} developed analytic hessian projection in a reduced basis setting but only applies to 2D planar cases. Energy specific approaches~\cite{McAdams:2011:EEC,Smith:2018:SNF} are able to reveal the analytic eigen system of certain distortion measurement, which are recently generalized by Smith et al.~\shortcite{Smith:2018:AEF} to cover more isotropic energies representable in stretch tensor invariants. Our work, as a further improvement over existing approaches, provides closed-form expressions for eigen values and eigen vectors of distortion energies parameterized by principal stretches, which is the most general and convenient way to define distortion measurement in both geometry and simulation areas.

\section{Background}
\begin{figure}[h]
\centering
\includegraphics[width=\linewidth]{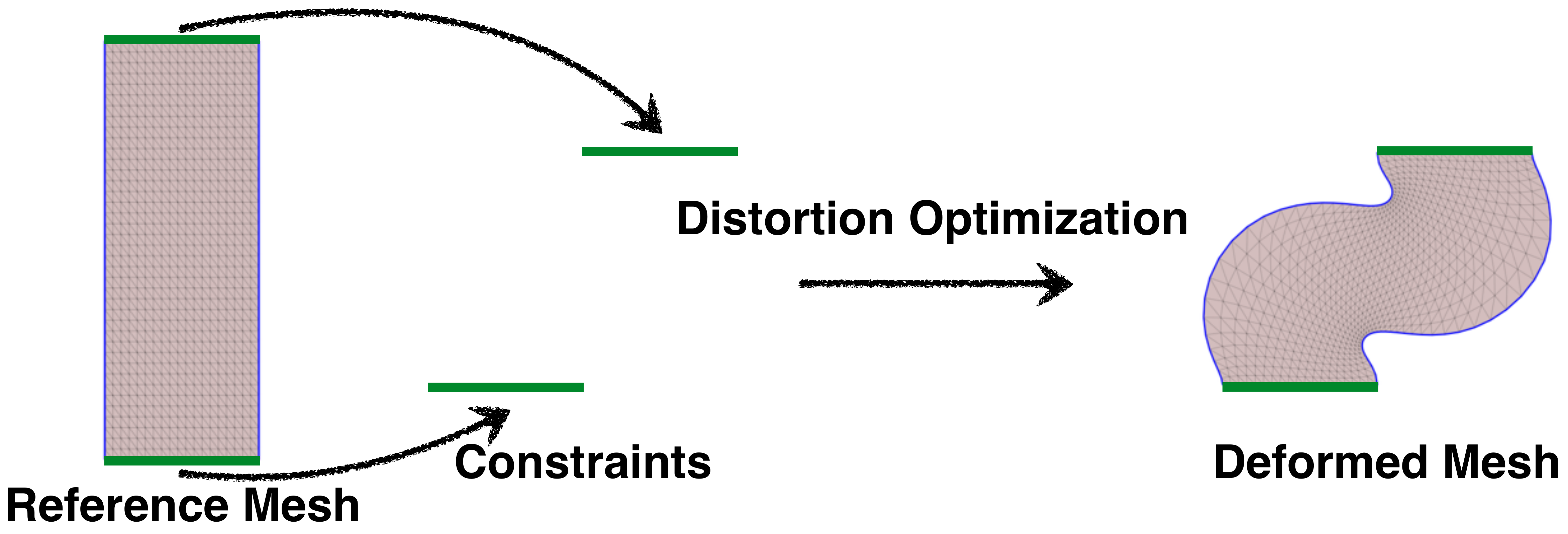}
\caption{}
\label{fig:illustration_example}
\end{figure}

Mesh distortion optimization is a computational approach to find mapping relationship between two domains, reference and deformed, in 2D or 3D. It has wide range of useful and essential applications in computer graphics field, for instance, planar animation, UV parameterization, geometry modeling, elastoplastic simulation, variational shape interpolation, etc. The goal is to preserve local metrics as much as possible with respect to constraints if any. Such problem starts from meshes, $\rest$, that discretize the reference domain and looks for optimal deformed meshes, $\deformed$, satisfying given constraints. (See \autoref{fig:illustration_example} as an illustration example.) Optimality judgement depends on the local metrics to be preserved.
One practical and popular direction to solve this problem is through formulating it as a constrained nonlinear local optimization problem,
\begin{equation}
\begin{split}
\min_{\deformed}&\hspace{1mm}\energy_{\rest}(\deformed)\\
\textbf{s.t.}&\hspace{2mm}\deformed\in\feasible,
\end{split}
\end{equation}
where $\feasible$ denotes the feasible set or region. Applying traditional constrained optimization techniques directly, like proximal gradient, barrier method, primal dual interior point method, etc, either suffers from slow convergence rate or stability issue caused by problem indefiness. In recent years, it attracts lots of attention and interests from researchers who proposed various efficient and stable solvers for this problem, including preconditioned gradient descent
\cite{Bouaziz:2014:PDF,Kovalsky:2016:AQP,Rabinovich:2017:SLI,Claici:2017:IPM,Liu:2018:PP}, variants of quasi-Newton
\cite{Liu:2017:QMR,Zhu:2018:BCQ,Peng:2018:AAG} and fixed Newton methods
\cite{Teran:2005:RQF,Stomakhin:2012:ECI,Shtengel:2017:GOV,Smith:2018:SNF,Smith:2018:AEF}. Among these approaches, fixed Newton methods demonstrate the most promising second order convergence rate property, especially when close to the local optimum. Like traditional Newton method, such approaches iteratively look for the solution where the mesh distortion energy gradient, $\nabla\energy_{\rest}(\deformed)$, vanishes. During each iteration, a crucial step is to solve a linear system for the solution update direction,
\begin{equation}
\systemhessian \search = -\nabla\energy_{\rest},
\end{equation}
where $\systemhessian$ represents energy hessian and $\search$ is the update direction. For monotonic energy descent approaches, $\systemhessian$ is required to be symmetric positive definite (SPD) in order to guarantee converged iterations. However, for general nonlinear energy, like $\energy_{\rest}$, its hessian usually doesn't hold such property everywhere within the feasible region $\feasible$. It is well known that the indefiness or negative definess of $\systemhessian$ will cause the Newton iteration diverge. In order to stabilize the solver, fixed Newton methods first project $\systemhessian$ onto SPD cone and then use its SPD projection, $\systemhessian^{\textbf{+}}$, to solve for update direction,
\begin{equation}
\systemhessian^{\textbf{+}} \search = -\nabla\energy_{\rest}.
\end{equation}
Composite majorization\cite{Shtengel:2017:GOV} obtains the projection through convex-concave decomposition of $\energy_{\rest}$ and uses the convex part's hessian as $\systemhessian^{\textbf{+}}$, but so far efficient decomposition method is limited to 2D setting only. Projected Newton methods adopt the more general eigen decomposition approach,
\begin{equation}\label{eqn:spd_projection}
\systemhessian = \projection^T\diagonal\projection\hspace{2mm}\Rightarrow\hspace{2mm}\systemhessian^{\textbf{+}} = \projection^T\diagonal^{\textbf{+}}\projection,
\end{equation}
by clamping all negative and nearly zero eigenvalues of $\diagonal$ to a small positive threshold, where $\diagonal^{\textbf{+}}$ is the clamped diagonal eigenvalue matrix. However, cost of eigen decomposition grows very quickly as the system size increases. To overcome this problem, researchers \cite{Teran:2005:RQF,Stomakhin:2012:ECI} explored and found that they can first efficiently convert $\systemhessian$ into a block diagonal form through congruent transformation and then only apply eigen decomposition to small sub-blocks, which is much faster than applying to $\systemhessian$ directly. Recently, Smith et al. \shortcite{Smith:2018:SNF,Smith:2018:AEF} further improve this result by assuming the distortion energy, $\energy_{\rest}$, is representable using invariants of stretch tensor. In such cases, they show that at least $\frac{2}{3}$ portion in 3D and $\frac{1}{2}$ portion in 2D of eigenvalues have analytical expressions. Even though such assumption already covers many popular energy choices used in geometry and physical simulation problems, it still restricts its application to limited range of distortion energy considering energy that can be defined by more atomic representation, principal stretches. We are showing in this work that for more general energy defined using principal stretches, the same portion of eigenvalues can always be evaluated analytically. Whether analytic expressions exist for the left portion of eigenvalues depends on the energy definition in terms of principal stretches and can be told by studying a small matrix, 3$\times$3 in 3D and 2$\times$2 in 2D.

\subsection{Problem Definition}

For the ease of explaination, we assume the reference domain is discretized using P1 element for 2D and 3D problems and our derivation should be generalizable to other types of elements, like P2, Q1, S1, etc. Moreover, in this paper, we only discuss 3D case and it's straightforward to extend our analysis to 2D case. Our input would be a 3D reference domain discretized using conforming tetrahedra mesh with piecewise linear shape functions. The function space spanned by these piecewise linear basis is supposed to contain an approximate domain deformation solution, whose approximation error should be orthogonal to this function space. We use $\rest_i$ to represent position of reference mesh's vertex $i$ and $\deformed_i$ to represent its corresponding position in deformed mesh. Each position vector will have three entries in 3D, for example, $\deformed_i = [\deformed_{i0}, \deformed_{i1}, \deformed_{i2}]^T$, representing its $x$-, $y$- and $z$- coordinates. In order to find optimal deformed mesh with minimal distortion with respect to reference $\rest$, we need to define distortion energy, $\energy_{\rest}$, that measures the deformation quality. A common choice adopted in computer graphics community to define $\energy_{\rest}$ is by aggregating distortion of each mesh element
\begin{equation}\label{eqn:distortion_energy}
\energy_{\rest} = \sum_t \volume_t \kernel_{\rest}(\deformed),
\end{equation} 
where $\volume_t$ is the volume of $t$-th tetrahedra in reference mesh $\rest$ and $\kernel_{\rest}(\cdot)$ is the distortion kernel that measures a single element's deformation. It's clear that due to linearity of differential operator, energy hessian $\systemhessian$ would be linear combination of mesh elements' kernel hessian,
\begin{equation}
\systemhessian = \sum_t \volume_t \tilde{\systemhessian}_t,
\end{equation}
which also implies that we can enforce kernel hessian to be SPD instead of enforcing energy hessian directly. The benefit is that we can largely reduce the problem size of the SPD projection step. As distortion kernel, $\kernel_{\rest}$, is defined per mesh element, which only involves four vertices of a tetrahedra, $\tilde{\systemhessian}$ has a low rank decomposition,
\begin{equation}
\tilde{\systemhessian} = \textbf{P}^T\hessian\textbf{P},
\end{equation}
where $\textbf{P} \in \mathbb{R}^{12\times 3n}$ is a permutation matrix for mesh with $n$ vertices and $\hessian \in \mathbb{R}^{12\times 12}$ is the second order differential of $\kernel_{\rest}$ with respect to single tetrahedra element's 12 vertex coordinates. Projected Newton methods then explore various approaches to efficiently diagonalize $\hessian$ in order to apply SPD projection as in \autoref{eqn:spd_projection}. As this projection need to be applied for every mesh element during each iteration, its computational cost will severely affect the solver's performance. Conventional eigen decomposition method relies on two-step algorithm, Householder reflection and shifted QR iteration, whose efficiency is not acceptable for per element computation. Existing projected Newton methods either rely on numerical eigenvalue computation or only support limited distortion energy. In this work, we present an analytic eigen system for the most general representation of distortion energy parameterized using principal stretches. For distortion energy that is representable using invariants of stretch tensor, it's not surprising to see that our results match exactly what have been shown by Smith et al.~\shortcite{Smith:2018:AEF}. However, our approach is much more flexible and applicable to a much broader class of distortion energy.

As we are interested in how to diagonalize per element $\hessian$ efficiently, it's enough to focus on single tetrahedar case. To simplify math notation, we will use $\rest_i$ and $\deformed_i$, $i \in \{0, 1, 2, 3\}$ to represent the element's four vertices in reference and deformed mesh. Moreover, if we assume the reference mesh never changes, which is usually the assumption adopted by geometry optimization and hyperelastic simulation, we can drop the subscript and use $\kernel$ to represent distortion kernel. For situations where reference mesh does change, like plastic deformation, our derivation can be extended very easily.

\subsection{Energy Parameterization}

An essential property of $\kernel$ is rigid motion invariant, which means it remains constant when rotating or translating the deformed mesh. For isotropic energy, such invariant property also holds when reference mesh is under rigid motion. There are quite a few ways to define $\kernel$ to satisfy this requirement, among which the most atomic, general and popular approach in both geometry and simulation fields is using principal stretch $\stretch$. $\stretch$ is defined as the singular value of deformation gradient $\dg = \frac{\partial \deformed}{\partial \rest}$. For piecewise linear shape function, it can be evaluated as
\begin{equation}\label{eqn:deformation_gradient}
\begin{split}
&\dg = \textbf{D}_w\textbf{D}_m^{-1},\\
\textbf{D}_w =& \begin{bmatrix}  \deformed_1 - \deformed_0, & \deformed_2 - \deformed_0, & \deformed_3 - \deformed_0\end{bmatrix},\\
\textbf{D}_m =& \begin{bmatrix}  \rest_1 - \rest_0, & \rest_2 - \rest_0, & \rest_3 - \rest_0\end{bmatrix}.
\end{split}
\end{equation}
And the principal strecthes are evaluated by its singular value decompostion (SVD),
\begin{equation}\label{eqn:svd_formula}
\begin{split}
&\dg = \textbf{U}\boldsymbol{\Sigma}\textbf{V}^T,\\
\boldsymbol{\Sigma} =& \begin{bmatrix} \stretch_0 & & \\ & \stretch_1 & \\ & & \stretch_2 \end{bmatrix},
\end{split}
\end{equation}
For energy that accepts invertible configuration, we adopt signed SVD where $\textbf{U}$ and $\textbf{V}$ are always rotation matrix. Thus we can parameterize distortion kernel through $\stretch$ as
\begin{equation}\label{eqn:kernel_parameterization}
\kernel(\deformed) \coloneqq \kernel(\stretch_0(\deformed), \stretch_1(\deformed), \stretch_2(\deformed)).
\end{equation}
The first order differential of $\kernel$ can then be evaluated as
\begin{equation}\label{eqn:kernel_gradient}
\nabla\kernel = \begin{bmatrix} \frac{\partial \kernel}{\partial \deformed_{00}} \\ \frac{\partial \kernel}{\partial \deformed_{01}} \\ \frac{\partial \kernel}{\partial \deformed_{02}} \\ \vdots \\  \frac{\partial \kernel}{\partial \deformed_{30}} \\ \frac{\partial \kernel}{\partial \deformed_{31}} \\ \frac{\partial \kernel}{\partial \deformed_{32}}\end{bmatrix}, \frac{\partial \kernel}{\partial \deformed_{ij}} = \sum_{k = 0}^2 \frac{\partial \kernel}{\partial \stretch_k}\frac{\partial \stretch_k}{\partial \deformed_{ij}},
\end{equation}
and its second order differential $\hessian$ can be written as
\begin{equation}\label{eqn:hessian_matrix}
\hessian = \begin{bmatrix} \frac{\partial^2 \kernel}{\partial \deformed_{00} \partial \deformed_{00}} & \cdots & \frac{\partial^2 \kernel}{\partial \deformed_{00} \partial \deformed_{32}}\\
\vdots & \ddots & \vdots\\
\frac{\partial^2 \kernel}{\partial \deformed_{32} \partial \deformed_{00}} & \cdots & \frac{\partial^2 \kernel}{\partial \deformed_{32} \partial \deformed_{32}} \end{bmatrix},
\end{equation}
where each entry is defined as
\begin{equation}\label{eqn:hessian_entry}
\frac{\partial^2 \kernel}{\partial \deformed_{ip} \partial \deformed_{jq}} = \sum_{k = 0}^2 \sum_{l = 0}^2 \frac{\partial^2\kernel}{\partial \stretch_k \stretch_l}\frac{\partial\stretch_l}{\partial\deformed_{ip}}\frac{\partial\stretch_k}{\partial\deformed_{jq}} + \sum_{k = 0}^2 \frac{\partial \kernel}{\partial \stretch_k} \frac{\partial^2\stretch_k}{\partial\deformed_{ip}\partial\deformed_{jq}}.
\end{equation}
Other choices of parameterization are also possible, for example, invariants of stretch tensor \cite{Smith:2018:AEF}, $I_1 = \sum_i\stretch_i, I_2 = \sum_i\stretch_i^2, I_3 = \prod_i\stretch_i$, or invariants of Cauchy-Green strain tensor \cite{Teran:2005:RQF}, $\rom{1} = \sum_i\stretch_i^2, \rom{2} = \sum_{i \neq j}\stretch_i^2\stretch_j^2, \rom{3} = \prod_i\stretch_i$, which are all easily representable in principal stretches but not vice versa. Thus our work provides the most general analysis framework for eigen system of $\hessian$.

\subsection{SVD Differential}

Before diving into the detailed discussion, we introduce one more concept, SVD differential, which serves as an essential building block to our eigen analysis. Even though SVD computation (\autoref{eqn:svd_formula}) relies on iterative numerical algorithms, SVD differential provides analytical derivatives of SVD components as long as parameterization of $\dg$ is given. Here we just list several important results that will be used in our later discussion and leave all detailed derivations in \autoref{append:svd_differential}. Suppose $\dg$ is parameterized by some scalar $x$, then we have
\begin{equation}\label{eqn:svd_first_derivative}
\frac{\partial \stretch_i}{\partial x} = \textbf{U}_i^T\frac{\partial\dg}{\partial x}\textbf{V}_i, \textbf{U}^T\frac{\partial \textbf{U}}{\partial x} = \omega^u_{x}, \frac{\partial \textbf{V}^T}{\partial x}\textbf{V} = -\omega^v_{x},
\end{equation}
where both $\omega_x^u$ and $\omega_x^v$ are skew-symmetric matrix,
\begin{equation}\label{eqn:skew_matrix_layout}
\omega = \begin{bmatrix}0 & \omega^0 & \omega^1 \\ -\omega^0 & 0 & \omega^2 \\ -\omega^1 & -\omega^2 & 0\end{bmatrix}.
\end{equation}
Their entries can be computed by solving three 2$\times$2 linear systems, such as
\begin{equation}\label{eqn:skew_linear_system}
\begin{bmatrix}\stretch_1 & -\stretch_0 \\ -\stretch_0 & \stretch_1\end{bmatrix}\begin{bmatrix}\omega_x^{u0} \\ \omega_x^{v0} \end{bmatrix} = \begin{bmatrix} \textbf{U}^T_0\frac{\partial \dg}{\partial x}\textbf{V}_1 \\ \textbf{U}^T_1\frac{\partial \dg}{\partial x}\textbf{V}_0 \end{bmatrix}.
\end{equation}
Such systems will become singular when two principal stretches are identical or sum to zero, which is a numerical stability issue that requires special treatment as shown in previous work \cite{Stomakhin:2012:ECI,Sin:2011:IIH,Xu:2015:NMD}. Same as Smith et al.\cite{Smith:2018:AEF}, we avoid such annoying numerical artifact by providing the analytical eigen system directly. 
%In \autoref{append:numerical_stability}, we provide more analysis results on such ill conditioning topic. 
Finally, we still need the second order derivatives of principal stretches. Suppose now $\dg$ can be parameterized by $x$ and $y$, then we have
\begin{equation}\label{eqn:svd_hessian}
\frac{\partial^2 \boldsymbol{\Sigma}}{\partial x \partial y} = \textit{diag}(\omega_x^u\boldsymbol{\Sigma}\omega_y^v + \omega_y^u\boldsymbol{\Sigma}\omega_x^v - \boldsymbol{\Sigma}\omega_x^v\omega_y^v - \omega_y^u\omega_x^u\boldsymbol{\Sigma}),
\end{equation}
where $\textit{diag}(\cdot)$ extracts the diagonal part of the input matrix.

\section{Eigen Analysis}

In this section, we will show the derivation of finding the eigen system for distortion kernel's second order differential $\hessian$ defined in \autoref{eqn:hessian_matrix}. We demonstrate that for 3D problems, $\hessian$ has a null space of dimension three and six of its eigen pairs can be analytically obtained while whether the other three have analytical expressions depends on the kernel definition in terms of principal stretches. As an overview of our approach, we first explore the null space of $\hessian$ and reduce our problem from $\hessian\in\mathbb{R}^{12\times 12}$ to $\tilde{\hessian}\in\mathbb{R}^{9\times 9}$. Next, we show that the $\tilde{\hessian}$ can be decomposed into one 3$\times$3 block and one 6$\times$6 diagonal block that contains the eigen values with analytical expressions. Finally, we apply our results to several energy kernel examples to demonstrate its flexibility and usefulness. Again, we will only provide results of essential steps and encourage interested readers to appendix for detailed derivations.

\subsection{From $\mathbb{R}^{12\times 12}$ To $\mathbb{R}^{9\times 9}$}

We first show that $\hessian\in\mathbb{R}^{12\times 12}$ can be factorized as $\hessian = \projection^T\tilde{\hessian}\projection$, $\projection \in \mathbb{R}^{9\times12}$, $\tilde{\hessian} \in \mathbb{R}^{9\times9}$. This is due to the fact that $\hessian$ has a null space of dimension three. For 3D P1 element, we have the following equality always hold,
\begin{equation}\label{eqn:zero_net_force}
\frac{\partial \kernel}{\partial \deformed_0} + \frac{\partial \kernel}{\partial \deformed_1} + \frac{\partial \kernel}{\partial \deformed_2} + \frac{\partial \kernel}{\partial \deformed_3} \equiv \textbf{0}.
\end{equation}
Intuitively, this means kernel derivative acts passively. Or if we take a physical point of view, internal forces should cancel out when there is no external forces. Detailed proof is provided in \autoref{append:net_force}. An immediate result is that we can represent $\frac{\partial \kernel}{\partial \deformed_i}$ using the other three. For example, \autoref{eqn:kernel_gradient} can be rewritten as
\begin{equation}
\nabla\kernel = \begin{bmatrix} \frac{\partial \kernel}{\partial \deformed_{00}} \\ \frac{\partial \kernel}{\partial \deformed_{01}} \\ \frac{\partial \kernel}{\partial \deformed_{02}} \\ \vdots \\ -(\frac{\partial \kernel}{\partial \deformed_{00}} + \frac{\partial \kernel}{\partial \deformed_{10}} + \frac{\partial \kernel}{\partial \deformed_{20}})  \\ -(\frac{\partial \kernel}{\partial \deformed_{01}} + \frac{\partial \kernel}{\partial \deformed_{11}} + \frac{\partial \kernel}{\partial \deformed_{21}}) \\ -(\frac{\partial \kernel}{\partial \deformed_{02}} + \frac{\partial \kernel}{\partial \deformed_{12}} + \frac{\partial \kernel}{\partial \deformed_{22}})\end{bmatrix},
\end{equation}
where we replace the last three entries with linear combination of the first nine ones. Similar idea can also be applied to $\hessian$ which is derivative of $\nabla\kernel$. If we divide $\hessian$ into a block 2$\times$2 form,
\begin{equation}\label{eqn:12_to_9_hessian}
\begin{split}
\hessian = \begin{bmatrix}\tilde{\hessian} & \bar{\hessian} \\ \bar{\hessian}^T & \hat{\hessian}\end{bmatrix},
\hspace{2mm}
\tilde{\hessian} = \begin{bmatrix} \frac{\partial^2 \kernel}{\partial \deformed_{00} \partial \deformed_{00}} & \cdots & \frac{\partial^2 \kernel}{\partial \deformed_{00} \partial \deformed_{22}}\\
\vdots & \ddots & \vdots\\
\frac{\partial^2 \kernel}{\partial \deformed_{22} \partial \deformed_{00}} & \cdots & \frac{\partial^2 \kernel}{\partial \deformed_{22} \partial \deformed_{22}} \end{bmatrix},
\end{split}
\end{equation}
then entries of $\bar{\hessian}\in\mathbb{R}^{9\times 3}$ and $\hat{\hessian}\in\mathbb{R}^{3\times 3}$ can be represented as linear combination of entries in $\tilde{\hessian}\in\mathbb{R}^{9\times 9}$. Notice $\tilde{\hessian}$ is the second order differential of distortion kernel with respect to just the first three element vertex coordiantes, which excludes the fourth one's coordinates. As shown in \autoref{append:12_to_9}, we can thus have
\begin{equation}\label{eqn:12_to_9_projection}
\projection = \begin{bmatrix} \textbf{I} & & & -\textbf{I}\\ 
& \textbf{I} & & -\textbf{I}\\
& & \textbf{I} & -\textbf{I}\end{bmatrix}
\end{equation}
satisfying $\hessian = \projection^T\tilde{\hessian}\projection$. Here $\textbf{I}$ is $3\times 3$ identity matrix.

\subsection{Analytic Decomposition of $\tilde{\hessian}$}

Given the above factorization result, we only need to show how SPD projection can be efficiently applied to $\tilde{\hessian}$ in the following. Thus we are going to demonstrate how $\tilde{\hessian}$ can be analytically decomposed into one 3$\times$3 block and one 6$\times$6 diagonal block. According to \autoref{eqn:hessian_entry}, we can also separate $\tilde{\hessian}$ into two terms,
\begin{equation}\label{eqn:hessian_separation}
\tilde{\hessian} = \tilde{\hessian}^{\#} + \tilde{\hessian}^{*},
\end{equation}
where entries of $\tilde{\hessian}^{\#}$ and $\tilde{\hessian}^{*}$ are $\sum_{k = 0}^2 \sum_{l = 0}^2 \frac{\partial^2\kernel}{\partial \stretch_k \stretch_l}\frac{\partial\stretch_l}{\partial\deformed_{ip}}\frac{\partial\stretch_k}{\partial\deformed_{jq}}$ and $\sum_{k = 0}^2 \frac{\partial \kernel}{\partial \stretch_k}\frac{\partial^2\stretch_k}{\partial\deformed_{ip}\partial\deformed_{jq}}$, $i, j, p, q\in \{0, 1, 2\}$ correspondingly. It's easy to see that if both $\tilde{\hessian}^{\#}$ and $\tilde{\hessian}^{*}$ are SPD, $\tilde{\hessian}$ is also SPD. Again we are going to decompose them into some simpler forms where SPD projection is efficient to be applied. We will show that $\tilde{\hessian}^{\#}$ can be decomposed into a 3$\times$3 block that only depends on distortion kernel definition in terms of principal stretches while $\tilde{\hessian}^{*}$ can be decomposed into diagonal form. Based on entry formulation of $\tilde{\hessian}^{\#}$, it's easy to see that it has the following decomposition, $\tilde{\hessian}^{\#} = (\tilde{\projection}^{\#})^T\tilde{\eigen}^{\#}\tilde{\projection}^{\#}$, where
\begin{equation}
\tilde{\textbf{D}}^{\#} = \begin{bmatrix} 	
\frac{\partial^2\kernel}{\partial\stretch_0\partial\stretch_0} & \frac{\partial^2\kernel}{\partial\stretch_0\partial\stretch_1} & \frac{\partial^2\kernel}{\partial\stretch_0\partial\stretch_2}\\
\frac{\partial^2\kernel}{\partial\stretch_1\partial\stretch_0} & \frac{\partial^2\kernel}{\partial\stretch_1\partial\stretch_1} & \frac{\partial^2\kernel}{\partial\stretch_1\partial\stretch_2}\\
\frac{\partial^2\kernel}{\partial\stretch_2\partial\stretch_0} & \frac{\partial^2\kernel}{\partial\stretch_2\partial\stretch_1} & \frac{\partial^2\kernel}{\partial\stretch_2\partial\stretch_2}
\end{bmatrix},
\hspace{1mm}
\tilde{\projection}^{\#} = \begin{bmatrix}	
\frac{\partial \stretch_0}{\partial \deformed_{00}} & \cdots & \frac{\partial \stretch_0}{\partial \deformed_{22}} \\ 
\frac{\partial \stretch_1}{\partial \deformed_{00}} & \cdots & \frac{\partial \stretch_1}{\partial \deformed_{22}} \\ 
\frac{\partial \stretch_2}{\partial \deformed_{00}} & \cdots & \frac{\partial \stretch_2}{\partial \deformed_{22}}
\end{bmatrix}.
\end{equation}
As $\tilde{\textbf{D}}^{\#}$ is the second order differential of distortion kernel with respect to principal stretches, it only requires kernel definition, which is usually provided before hand, to determine its diagonalizability. In \autoref{}, we will utilize this decomposition to provide analytical eigen pairs for several distortion kernel examples. In cases where no analytical eigen expressions exist for this 3$\times$3 system, we adopt numerical solutions instead.

For $\tilde{\hessian}^{*}$, it's not easy to see similar decomposition immediately. We leave all detailed derivations in \autoref{append:diagonalize_6_by_6} and give the essential steps here. Similar to \autoref{eqn:hessian_separation}, we first separate $\tilde{\hessian}^{*}$ into three terms,
\begin{equation}\label{eqn:three_term_summation_separation}
\tilde{\hessian}^{*} = \tilde{\hessian}^{*_{0, 1}} + \tilde{\hessian}^{*_{1, 2}} + \tilde{\hessian}^{*_{2, 0}},
\end{equation}
where each term will have a decomposition. Next, we take $\tilde{\hessian}^{*_{0, 1}}$ as an illustration example and the other two can be decomposed in similar manner. As a first step, we obtain the following decomposition,
\begin{equation}\label{eqn:temp_decomposition}
\begin{split}
\tilde{\hessian}^{*_{0, 1}} &= (\tilde{\projection}^{\star_{0, 1}})^T \tilde{\eigen}^{\star_{0, 1}} \tilde{\projection}^{\star_{0, 1}},
\hspace{2mm}
\tilde{\projection}^{\star_{0, 1}} = \begin{bmatrix}
\omega^{u0}_{\deformed_{00}} & \cdots & \omega^{u0}_{\deformed_{22}}\\
\omega^{v0}_{\deformed_{00}} & \cdots & \omega^{v0}_{\deformed_{22}}
\end{bmatrix},\\
&\tilde{\eigen}^{\star_{0, 1}} = \begin{bmatrix}
\stretch_0\frac{\partial\kernel}{\partial\stretch_0} + \stretch_1\frac{\partial\kernel}{\partial\stretch_1} & -(\stretch_0\frac{\partial\kernel}{\partial\stretch_1} + \stretch_1\frac{\partial\kernel}{\partial\stretch_0}) \\
-(\stretch_1\frac{\partial\kernel}{\partial\stretch_0} + \stretch_0\frac{\partial\kernel}{\partial\stretch_1}) & \stretch_1\frac{\partial\kernel}{\partial\stretch_1} + \stretch_0\frac{\partial\kernel}{\partial\stretch_0}
\end{bmatrix},
\end{split}
\end{equation}
where each column vector of $\tilde{\projection}^{\star_{0, 1}}\in\mathbb{R}^{2\times 9}$ can be obtained by solving a small 2$\times$2 linear system as shown in \autoref{eqn:skew_linear_system}. Thus we have
\begin{equation}
\begin{bmatrix} \omega_{\deformed_{ip}}^{u0} \\ \omega_{\deformed_{ip}}^{v0} \end{bmatrix} = \frac{1}{\stretch_1^2 - \stretch_0^2}\begin{bmatrix} \stretch_1 & \stretch_0 \\ \stretch_0 & \stretch_1 \end{bmatrix}\begin{bmatrix} \textbf{U}^T_0\frac{\partial \dg}{\partial \deformed_{ip}}\textbf{V}_1 \\ \textbf{U}^T_1\frac{\partial \dg}{\partial \deformed_{ip}}\textbf{V}_0 \end{bmatrix},\hspace{2mm}i, p\in \{0, 1, 2\}.
\end{equation}
By regrouping matrix products, we obtain a new decomposition for $\tilde{\hessian}^{*_{0, 1}}$ as $(\tilde{\projection}^{\bullet_{0, 1}})^T \tilde{\eigen}^{\bullet_{0, 1}} \tilde{\projection}^{\bullet_{0, 1}}$, where
\begin{equation}
\begin{split}
\tilde{\eigen}^{\bullet_{0, 1}} &=  \frac{1}{(\stretch_1^2 - \stretch_0^2)^2}\begin{bmatrix} \stretch_1 & \stretch_0 \\ \stretch_0 & \stretch_1 \end{bmatrix}^T
\tilde{\eigen}^{\star_{0, 1}}
\begin{bmatrix} \stretch_1 & \stretch_0 \\ \stretch_0 & \stretch_1 \end{bmatrix},\\
&\tilde{\projection}^{\bullet_{0, 1}} = \begin{bmatrix}
\textbf{U}^T_0\frac{\partial \dg}{\partial \deformed_{00}}\textbf{V}_1 & \cdots & \textbf{U}^T_0\frac{\partial \dg}{\partial \deformed_{22}}\textbf{V}_1\\
\textbf{U}^T_1\frac{\partial \dg}{\partial \deformed_{00}}\textbf{V}_0 & \cdots & \textbf{U}^T_1\frac{\partial \dg}{\partial \deformed_{22}}\textbf{V}_0
\end{bmatrix}.
\end{split}
\end{equation}
In this case, we can diagonalize $\tilde{\eigen}^{\bullet_{0, 1}}$ as
\begin{equation}\label{eqn:2x2_diagonalization}
\tilde{\eigen}^{\bullet_{0, 1}} = \begin{bmatrix}-1 & 1 \\ 1 & 1\end{bmatrix}^T
\begin{bmatrix}\frac{\frac{\partial\kernel}{\partial\stretch_0} - \frac{\partial\kernel}{\partial\stretch_1}}{\stretch_0 - \stretch_1} & 0 \\ 0 & \frac{\frac{\partial\kernel}{\partial\stretch_0} + \frac{\partial\kernel}{\partial\stretch_1}}{\stretch_0 + \stretch_1}\end{bmatrix}
\begin{bmatrix}-1 & 1 \\ 1 & 1\end{bmatrix}.
\end{equation}
To see this more clearly, we first separate $\tilde{\eigen}^{\bullet_{0, 1}}$ into sum of the following two terms,
\begin{equation}
\tilde{\eigen}^{\bullet_{0, 1}} = \frac{\frac{\partial\kernel}{\partial\stretch_0} - \frac{\partial\kernel}{\partial\stretch_1}}{2(\stretch_0 - \stretch_1)}\begin{bmatrix}1 & -1 \\ -1 & 1\end{bmatrix} + \frac{\frac{\partial\kernel}{\partial\stretch_0} + \frac{\partial\kernel}{\partial\stretch_1}}{2(\stretch_0 + \stretch_1)}\begin{bmatrix}1 & 1 \\ 1 & 1\end{bmatrix},
\end{equation}
where the two constant 2$\times$2 matrices have simple diagonal forms as
\begin{equation}
\begin{split}
\begin{bmatrix}1 & -1 \\ -1 & 1\end{bmatrix} &= \begin{bmatrix}-1 & 1 \\ 1 & 1\end{bmatrix}^T
\begin{bmatrix}2 & 0 \\ 0 & 0\end{bmatrix}
\begin{bmatrix}-1 & 1 \\ 1 & 1\end{bmatrix},\\
\begin{bmatrix}1 & 1 \\ 1 & 1\end{bmatrix} &= \begin{bmatrix}-1 & 1 \\ 1 & 1\end{bmatrix}^T
\begin{bmatrix}0 & 0 \\ 0 & 2\end{bmatrix}
\begin{bmatrix}-1 & 1 \\ 1 & 1\end{bmatrix}.
\end{split}
\end{equation}
Notice both matrices share the same eigen space, so we can combine these two factorizations together and obtain \autoref{eqn:2x2_diagonalization}. Finally, given the above results, we can diagonalize $\tilde{\hessian}^{*_{0, 1}}$ through decomposition, $(\tilde{\projection}^{*_{0, 1}})^T \tilde{\eigen}^{*_{0, 1}} \tilde{\projection}^{*_{0, 1}}$, where
\begin{equation}\label{eqn:analytic_eigen_value}
\tilde{\eigen}^{*_{0, 1}} = \begin{bmatrix}\frac{\frac{\partial\kernel}{\partial\stretch_0} - \frac{\partial\kernel}{\partial\stretch_1}}{\stretch_0 - \stretch_1} & 0 \\ 0 & \frac{\frac{\partial\kernel}{\partial\stretch_0} + \frac{\partial\kernel}{\partial\stretch_1}}{\stretch_0 + \stretch_1}\end{bmatrix}, 
\hspace{2mm}
\tilde{\projection}^{*_{0, 1}} = \begin{bmatrix}-1 & 1 \\ 1 & 1\end{bmatrix} \tilde{\projection}^{\bullet_{0, 1}}.
\end{equation}
For the other two matrices, $\tilde{\hessian}^{*_{1, 2}}$ and $\tilde{\hessian}^{*_{2, 0}}$, we can obtain similar results, which justifies our claim that six eigen pairs of $\hessian$ have analytical forms. If we combine all the derivations from this section together, we obtain the analytical decomposition of $\hessian$ as 
\begin{equation}
\begin{split}
&\hessian = \projection^T\tilde{\hessian}\projection = \projection^T\tilde{\projection}^T\tilde{\eigen}\tilde{\projection}\projection,\\
\tilde{\eigen} =& 
\begin{bmatrix} 
\tilde{\eigen}^{\#} & & &\\
& \tilde{\eigen}^{*_{0, 1}} & &\\
& & \tilde{\eigen}^{*_{1, 2}} &\\
& & & \tilde{\eigen}^{*_{2, 0}}
\end{bmatrix},
\tilde{\projection} =
\begin{bmatrix}
\tilde{\projection}^{\#} \\ \tilde{\projection}^{*_{0, 1}} \\ \tilde{\projection}^{*_{1, 2}} \\ \tilde{\projection}^{*_{2, 0}}
\end{bmatrix}.
\end{split}
\end{equation}
Moreover, the six eigen values with analytical expressions may cause numerical instability issues as sum and difference of principal stretches appear as denominators (see \autoref{eqn:2x2_diagonalization}). At the first glance, such annoying problem seems to be unavoidable but depends on the distortion kernel definition. 
%In \autoref{append:numerical_stability}, we will give a detailed analysis and show that at least half of the issues are avoidable and the other half relies on type of energy kernel being used. 

\bibliographystyle{ACM-Reference-Format}
\bibliography{paper}

\appendix

\section{SVD Differential}
\label{append:svd_differential}

We provide detailed derivations to compute SVD differential as shown in \autoref{eqn:svd_first_derivative} and \autoref{eqn:svd_hessian}. If $\dg$ is parameterized by $x$ and according to SVD factorization as well as product rule, we can have the following result by taking derivative with respect to $x$ on both sides of \autoref{eqn:svd_formula},
\begin{equation}\label{eqn:svd_gradient}
\begin{split}
&\frac{\partial \dg}{\partial x} = \frac{\partial \textbf{U}}{\partial x}\boldsymbol{\Sigma}\textbf{V}^T + \textbf{U}\frac{\partial \boldsymbol{\Sigma}}{\partial x}\textbf{V}^T + \textbf{U}\boldsymbol{\Sigma}\frac{\partial \textbf{V}^T}{\partial x}\\
\Rightarrow& \textbf{U}^T\frac{\partial \dg}{\partial x}\textbf{V} = \textbf{U}^T\frac{\partial \textbf{U}}{\partial x}\boldsymbol{\Sigma} + \frac{\partial \boldsymbol{\Sigma}}{\partial x} + \boldsymbol{\Sigma}\frac{\partial \textbf{V}^T}{\partial x}\textbf{V}.
\end{split}
\end{equation}
If we already know $\dg$ in terms of $x$, then we can compute rotation differential $\frac{\partial \textbf{U}}{\partial x}$, $\frac{\partial \textbf{V}^T}{\partial x}$ and singular value differential $\frac{\partial \boldsymbol{\Sigma}}{\partial x}$ given the fact that $\textbf{U}^T\frac{\partial \textbf{U}}{\partial x}$, $\frac{\partial \textbf{V}^T}{\partial x}\textbf{V}$ are skew-symmetric and $\frac{\partial \boldsymbol{\Sigma}}{\partial x}$ is diagonal. This is also known as SVD gradient, which has been investigated before \cite{Papadopoulo:2000:EJS}. To compute first order derivative of distortion kernel, $\nabla \kernel$, this is already enough as we only need to know $\frac{\partial \boldsymbol{\Sigma}}{\partial x}$. In order to compute second order derivative, $\hessian$, we also need to evaluate $\textbf{U}^T\frac{\partial \textbf{U}}{\partial x}$, $\frac{\partial \textbf{V}^T}{\partial x}\textbf{V}$ and the second order derivative of singular values. If now $\dg$ is parameterized by $x$ and $y$, we can apply similar trick as \autoref{eqn:svd_gradient},
\begin{equation}\label{eqn:svd_second_derivative}
\begin{split}
\frac{\partial^2 \dg}{\partial x \partial y} =& \frac{\partial^2 \textbf{U}}{\partial x \partial y}\boldsymbol{\Sigma}\textbf{V}^T + \frac{\partial \textbf{U}}{\partial x}\frac{\partial \boldsymbol{\Sigma}}{\partial y}\textbf{V}^T + \frac{\partial \textbf{U}}{\partial x}\boldsymbol{\Sigma}\frac{\partial\textbf{V}^T}{\partial y}\\
+& \frac{\partial \textbf{U}}{\partial y}\frac{\partial \boldsymbol{\Sigma}}{\partial x}\textbf{V}^T + \textbf{U}\frac{\partial^2 \boldsymbol{\Sigma}}{\partial x \partial y}\textbf{V}^T + \textbf{U}\frac{\partial \boldsymbol{\Sigma}}{\partial x}\frac{\partial \textbf{V}^T}{\partial y}\\
+& \frac{\partial \textbf{U}}{\partial y}\boldsymbol{\Sigma}\frac{\partial \textbf{V}^T}{\partial x} + \textbf{U}\frac{\partial \boldsymbol{\Sigma}}{\partial y}\frac{\partial \textbf{V}^T}{\partial x} + \textbf{U}\boldsymbol{\Sigma}\frac{\partial^2 \textbf{V}^T}{\partial x \partial y},\\
\Rightarrow \textbf{U}^T\frac{\partial^2 \dg}{\partial x \partial y}\textbf{V} =& \textbf{U}^T\frac{\partial^2 \textbf{U}}{\partial x \partial y}\boldsymbol{\Sigma} + \textbf{U}^T\frac{\partial \textbf{U}}{\partial x}\frac{\partial \boldsymbol{\Sigma}}{\partial y} + \textbf{U}^T\frac{\partial \textbf{U}}{\partial x}\boldsymbol{\Sigma}\frac{\partial\textbf{V}^T}{\partial y}\textbf{V}\\
+& \textbf{U}^T\frac{\partial \textbf{U}}{\partial y}\frac{\partial \boldsymbol{\Sigma}}{\partial x} + \frac{\partial^2 \boldsymbol{\Sigma}}{\partial x \partial y} + \frac{\partial \boldsymbol{\Sigma}}{\partial x}\frac{\partial \textbf{V}^T}{\partial y}\textbf{V}\\
+& \textbf{U}^T\frac{\partial \textbf{U}}{\partial y}\boldsymbol{\Sigma}\frac{\partial \textbf{V}^T}{\partial x}\textbf{V} + \frac{\partial \boldsymbol{\Sigma}}{\partial y}\frac{\partial \textbf{V}^T}{\partial x}\textbf{V} + \boldsymbol{\Sigma}\frac{\partial^2 \textbf{V}^T}{\partial x \partial y}\textbf{V}.
\end{split}
\end{equation}
Here we are interested in $\frac{\partial^2 \boldsymbol{\Sigma}}{\partial x \partial y}$, but the above equations also include other unknowns, $\frac{\partial^2 \textbf{U}}{\partial x \partial y}$ and $\frac{\partial^2 \textbf{V}^T}{\partial x \partial y}$. To get rid of them, we first spend some effort to see what they look like. Suppose we have a rotation matrix $\textbf{Q}$ with parameterization $x$ and $y$. Then we can derive the following result,
\begin{equation}
\begin{split}
&\textbf{Q}^T\textbf{Q} = \textbf{I} \Rightarrow \frac{\partial \textbf{Q}^T}{\partial x}\textbf{Q} + \textbf{Q}^T\frac{\partial \textbf{Q}}{\partial x} = \textbf{O},\\
\Rightarrow& \frac{\partial^2 \textbf{Q}^T}{\partial x \partial y}\textbf{Q} + \frac{\partial \textbf{Q}^T}{\partial x}\frac{\partial \textbf{Q}}{\partial y} + \frac{\partial \textbf{Q}^T}{\partial y}\frac{\partial \textbf{Q}}{\partial x} + \textbf{Q}^T\frac{\partial^2 \textbf{Q}}{\partial x \partial y} = \textbf{O},\\
\Rightarrow& \frac{\partial^2 \textbf{Q}^T}{\partial x \partial y}\textbf{Q} + \frac{\partial \textbf{Q}^T}{\partial x}\textbf{Q}\textbf{Q}^T\frac{\partial \textbf{Q}}{\partial y} + \frac{\partial \textbf{Q}^T}{\partial y}\textbf{Q}\textbf{Q}^T\frac{\partial \textbf{Q}}{\partial x} + \textbf{Q}^T\frac{\partial^2 \textbf{Q}}{\partial x \partial y} = \textbf{O},\\
\Rightarrow& \frac{\partial^2 \textbf{Q}^T}{\partial x \partial y}\textbf{Q} + \frac{\partial \textbf{Q}^T}{\partial x}\textbf{Q}\textbf{Q}^T\frac{\partial \textbf{Q}}{\partial y} = -(\textbf{Q}^T\frac{\partial^2 \textbf{Q}}{\partial x \partial y} + \frac{\partial \textbf{Q}^T}{\partial y}\textbf{Q}\textbf{Q}^T\frac{\partial \textbf{Q}}{\partial x})\\
&= -(\frac{\partial^2 \textbf{Q}^T}{\partial x \partial y}\textbf{Q} + \frac{\partial \textbf{Q}^T}{\partial x}\textbf{Q}\textbf{Q}^T\frac{\partial \textbf{Q}}{\partial y})^T.
\end{split}
\end{equation}
Thus $\textbf{U}^T\frac{\partial^2 \textbf{U}}{\partial x \partial y} + \frac{\partial \textbf{U}^T}{\partial y}\textbf{U}\textbf{U}^T\frac{\partial \textbf{U}}{\partial x}$ and $\frac{\partial^2 \textbf{V}^T}{\partial x \partial y}\textbf{V} + \frac{\partial \textbf{V}^T}{\partial x}\textbf{V}\textbf{V}^T\frac{\partial \textbf{V}}{\partial y}$ are also skew-symmetric. If we adopt the same math notation convention used in \autoref{eqn:svd_first_derivative}, we have
\begin{equation}
\begin{split}
&\omega_x^u \coloneqq \textbf{U}^T\frac{\partial \textbf{U}}{\partial x}, \omega_y^u \coloneqq \textbf{U}^T\frac{\partial \textbf{U}}{\partial y}, \omega_x^v \coloneqq -\frac{\partial \textbf{V}^T}{\partial x}\textbf{V}, \omega_y^v \coloneqq -\frac{\partial \textbf{V}^T}{\partial y}\textbf{V},\\
&\omega_{xy}^u \coloneqq \textbf{U}^T\frac{\partial^2 \textbf{U}}{\partial x \partial y} + \frac{\partial \textbf{U}^T}{\partial y}\textbf{U}\textbf{U}^T\frac{\partial \textbf{U}}{\partial x} = \textbf{U}^T\frac{\partial^2 \textbf{U}}{\partial x \partial y} - \omega_y^u\omega_x^u,\\
&\omega_{xy}^v \coloneqq -(\frac{\partial^2 \textbf{V}^T}{\partial x \partial y}\textbf{V} + \frac{\partial \textbf{V}^T}{\partial x}\textbf{V}\textbf{V}^T\frac{\partial \textbf{V}}{\partial y}) = -\frac{\partial^2 \textbf{V}^T}{\partial x \partial y}\textbf{V} + \omega_x^v\omega_y^v.
\end{split}
\end{equation}
Then we can rewrite \autoref{eqn:svd_second_derivative} as
\begin{equation}
\begin{split}
\textbf{U}^T\frac{\partial^2 \dg}{\partial x \partial y}\textbf{V} =& (\omega_{xy}^u + \omega_y^u\omega_x^u)\boldsymbol{\Sigma} + \omega_x^u\frac{\partial \boldsymbol{\Sigma}}{\partial y} - \omega_x^u\boldsymbol{\Sigma}\omega_y^v\\
+& \omega_y^u\frac{\partial \boldsymbol{\Sigma}}{\partial x} + \frac{\partial^2 \boldsymbol{\Sigma}}{\partial x \partial y} - \frac{\partial \boldsymbol{\Sigma}}{\partial x}\omega_y^v\\
-& \omega_y^u\boldsymbol{\Sigma}\omega_x^v - \frac{\partial \boldsymbol{\Sigma}}{\partial y}\omega_x^v - \boldsymbol{\Sigma}(\omega_{xy}^v - \omega_x^v\omega_y^v).
\end{split}
\end{equation}
Notice that product of skew-symmetric matrix and diagonal matrix has zero diagonals and the second order differential $\frac{\partial^2 \boldsymbol{\Sigma}}{\partial x \partial y}$ is diagonal matrix, we have
\begin{equation}
\frac{\partial^2 \boldsymbol{\Sigma}}{\partial x \partial y} = \textit{diag}(\omega_x^u\boldsymbol{\Sigma}\omega_y^v + \omega_y^u\boldsymbol{\Sigma}\omega_x^v - \boldsymbol{\Sigma}\omega_x^v\omega_y^v - \omega_y^u\omega_x^u\boldsymbol{\Sigma} + \textbf{U}^T\frac{\partial^2 \dg}{\partial x \partial y}\textbf{V}),
\end{equation}
where $\textit{diag}(\cdot)$ extracts the diagonal part of the input matrix. In our P1 element setting, $\dg$ is a linear function (see \autoref{eqn:deformation_gradient}), so its second order derivative is always zero. Thus we can further simplify our result and get \autoref{eqn:svd_hessian}.

\section{Zero Net Element Derivative}
\label{append:net_force}
We then provide proof for the statement of \autoref{eqn:zero_net_force}.
\begin{proof}
According to the definition of $\dg$ in \autoref{eqn:deformation_gradient}, it's easy to verify that the following equality always holds for any $p$-th coordinate,
\begin{equation}
\sum_{i = 0}^3\frac{\partial \dg}{\partial \deformed_{ip}} \equiv \textbf{O},\hspace{2mm}\forall p\in\{0, 1, 2\},
\end{equation}
where $\textbf{O}$ is a 3$\times$3 zero matrix. Then according to \autoref{eqn:svd_first_derivative}, we have 
\begin{equation}
\begin{split}
\sum_{i = 0}^3 \frac{\partial \stretch_k}{\partial \deformed_{ip}} = \textbf{U}_k^T(\sum_{i = 0}^3\frac{\partial \dg}{\partial \deformed_{ip}})\textbf{V}_k \equiv 0,\hspace{2mm}\forall k \in\{0, 1, 2\},\\
\Rightarrow \sum_{k = 0}^2\frac{\partial\kernel}{\partial\stretch_k}\sum_{i = 0}^3 \frac{\partial \stretch_k}{\partial \deformed_{ip}}\equiv 0 \Rightarrow \sum_{i = 0}^3\frac{\partial\kernel}{\partial\deformed_{ip}} \equiv 0,\hspace{2mm}\forall p\in\{0, 1, 2\}.
\end{split}
\end{equation}
As the above equality holds for any $p$-th coordinate, the statement of \autoref{eqn:zero_net_force} is always true.
\end{proof}

\section{From $\mathbb{R}^{12\times 12}$ To $\mathbb{R}^{9\times 9}$}
\label{append:12_to_9}

Next we utilize the result just proved to show $\hessian = \projection^T\tilde{\hessian}\projection$ as in \autoref{eqn:12_to_9_hessian} and \autoref{eqn:12_to_9_projection}. The key idea is that entries of $\bar{\hessian}$ and $\hat{\hessian}$ can be represented as linear combinations of entries from $\tilde{\hessian}$, where
\begin{equation}
\begin{split}
\bar{\hessian} = \begin{bmatrix} \frac{\partial^2 \kernel}{\partial \deformed_{00} \partial \deformed_{30}} & \cdots & \frac{\partial^2 \kernel}{\partial \deformed_{00} \partial \deformed_{32}}\\
\vdots & \ddots & \vdots\\
\frac{\partial^2 \kernel}{\partial \deformed_{22} \partial \deformed_{30}} & \cdots & \frac{\partial^2 \kernel}{\partial \deformed_{22} \partial \deformed_{32}} \end{bmatrix},\\
\hat{\hessian} = \begin{bmatrix} \frac{\partial^2 \kernel}{\partial \deformed_{30} \partial \deformed_{30}} & \cdots & \frac{\partial^2 \kernel}{\partial \deformed_{30} \partial \deformed_{32}}\\
\vdots & \ddots & \vdots\\
\frac{\partial^2 \kernel}{\partial \deformed_{32} \partial \deformed_{30}} & \cdots & \frac{\partial^2 \kernel}{\partial \deformed_{32} \partial \deformed_{32}} \end{bmatrix}.
\end{split}
\end{equation}
First, recall that \autoref{eqn:zero_net_force} holds true for every $p$-th coordinate and we have,
\begin{equation}\label{eqn:zero_net_force_entry}
\frac{\partial \kernel}{\partial \deformed_{0p}} + \frac{\partial \kernel}{\partial \deformed_{1p}} + \frac{\partial \kernel}{\partial \deformed_{2p}} + \frac{\partial \kernel}{\partial \deformed_{3p}} = 0,\hspace{2mm}\forall p\in\{0, 1, 2\}.
\end{equation}
To compute entries of $\bar{\hessian}$, namely $\frac{\partial^2 \kernel}{\partial \deformed_{jq} \partial \deformed_{3p}}$, $j, p, q \in \{0, 1, 2\}$, we simply take derivative of \autoref{eqn:zero_net_force_entry} with respect to $\deformed_{jq}$, $j, q\in\{0, 1, 2\}$ and have
\begin{equation}
-(\frac{\partial^2 \kernel}{\partial \deformed_{jq} \partial \deformed_{0p}} + \frac{\partial^2 \kernel}{\partial \deformed_{jq} \partial \deformed_{1p}} + \frac{\partial^2 \kernel}{\partial \deformed_{jq} \partial \deformed_{2p}}) = \frac{\partial^2 \kernel}{\partial \deformed_{jq} \partial \deformed_{3p}}.
\end{equation}
Then to compute entries of $\hat{\hessian}$, namely $\frac{\partial^2 \kernel}{\partial \deformed_{3p} \partial \deformed_{3q}}$, $p, q\in\{0, 1, 2\}$, we again take derivative of \autoref{eqn:zero_net_force_entry} with respect to $\deformed_{3q}$, $q\in\{0, 1, 2\}$ and have
\begin{equation}
-(\frac{\partial^2 \kernel}{\partial \deformed_{3q} \partial \deformed_{0p}} + \frac{\partial^2 \kernel}{\partial \deformed_{3q} \partial \deformed_{1p}} + \frac{\partial^2 \kernel}{\partial \deformed_{3q} \partial \deformed_{2p}}) = \frac{\partial^2 \kernel}{\partial \deformed_{3q} \partial \deformed_{3p}}.
\end{equation}
Thus we can represent every entry in $\bar{\hessian}$ and $\hat{\hessian}$ through linear combination of entries in $\tilde{\hessian}$. With little bit more effort, it's not hard to show the result of \autoref{eqn:12_to_9_projection}.

\section{Diagonalization of $\tilde{\hessian}^{*}$} 
\label{append:diagonalize_6_by_6}

In order to decompose $\tilde{\hessian}^{*}$ into diagonal form, we first expand its entries and regroup them as
\begin{equation}
\begin{split}
\frac{\partial^2 \stretch_0}{\partial x \partial y} = (\underbrace{\omega^{u0}_{x}\omega^{u0}_{y} + \omega^{v0}_{x}\omega^{v0}_{y}}_{S^{0,1}_{x,y}} + \underbrace{\omega^{u1}_{x}\omega^{u1}_{y} + \omega^{v1}_{x}\omega^{v1}_{y}}_{S^{2, 0}_{x,y}})\stretch_0\\
-(\underbrace{\omega^{u0}_{y}\omega^{v0}_{x} + \omega^{u0}_{x}\omega^{v0}_{y}}_{T^{0, 1}_{x,y}})\stretch_1 - (\underbrace{\omega^{u1}_{y}\omega^{v1}_{x} + \omega^{u1}_{x}\omega^{v1}_{y}}_{T^{2,0}_{x,y}})\stretch_2,\\
\frac{\partial^2 \stretch_1}{\partial x \partial y} = (\underbrace{\omega^{u0}_{x}\omega^{u0}_{y} + \omega^{v0}_{x}\omega^{v0}_{y}}_{S^{0, 1}_{x,y}} + \underbrace{\omega^{u2}_{x}\omega^{u2}_{y} + \omega^{v2}_{x}\omega^{v2}_{y}}_{S^{1, 2}_{x,y}})\stretch_1\\
-(\underbrace{\omega^{u0}_{y}\omega^{v0}_{x} + \omega^{u0}_{x}\omega^{v0}_{y}}_{T^{0, 1}_{x,y}})\stretch_0 - (\underbrace{\omega^{u2}_{y}\omega^{v2}_{x} + \omega^{u2}_{x}\omega^{v2}_{y}}_{T^{1, 2}_{x,y}})\stretch_2,\\
\frac{\partial^2 \stretch_2}{\partial x \partial y}= (\underbrace{\omega^{u2}_{x}\omega^{u2}_{y} + \omega^{v2}_{x}\omega^{v2}_{y}}_{S^{1, 2}_{x,y}} + \underbrace{\omega^{u1}_{x}\omega^{u1}_{y} + \omega^{v1}_{x}\omega^{v1}_{y}}_{S^{2, 0}_{x,y}})\stretch_2\\
-(\underbrace{\omega^{u2}_{y}\omega^{v2}_{x} + \omega^{u2}_{x}\omega^{v2}_{y}}_{T^{1, 2}_{x,y}})\stretch_1 - (\underbrace{\omega^{u1}_{y}\omega^{v1}_{x} + \omega^{u1}_{x}\omega^{v1}_{y}}_{T^{2, 0}_{x,y}})\stretch_0,
\end{split}
\end{equation}
where we adopt the same notation convention in \autoref{eqn:skew_matrix_layout}. Notice here we use $x$ and $y$ to represent vertex coordinates. Then for each entry of $\tilde{\hessian}^{*}$, we have
\begin{equation}
\begin{split}
\tilde{\textbf{H}}^{*}_{x, y} =[ (S^{0,1}_{x,y} + S^{2, 0}_{x,y})\stretch_0 - T^{0, 1}_{x,y}\stretch_1 - T^{2, 0}_{x,y}\stretch_2]\frac{\partial\kernel}{\partial\stretch_0}\\
+ [ (S^{0,1}_{x,y} + S^{1, 2}_{x,y})\stretch_1 - T^{0, 1}_{x,y}\stretch_0 - T^{1, 2}_{x,y}\stretch_2]\frac{\partial\kernel}{\partial\stretch_1}\\
+ [ (S^{1,2}_{x,y} + S^{2, 0}_{x,y})\stretch_2 - T^{1, 2}_{x,y}\stretch_1 - T^{2, 0}_{x,y}\stretch_0]\frac{\partial\kernel}{\partial\stretch_2}\\
= [ S^{0,1}_{x,y}\stretch_0 - T^{0, 1}_{x,y}\stretch_1]\frac{\partial\kernel}{\partial\stretch_0} +  [ S^{0,1}_{x,y}\stretch_1 - T^{0, 1}_{x,y}\stretch_0]\frac{\partial\kernel}{\partial\stretch_1}\\
+ [ S^{1, 2}_{x,y}\stretch_1 - T^{1, 2}_{x,y}\stretch_2]\frac{\partial\kernel}{\partial\stretch_1} + [ S^{1,2}_{x,y}\stretch_2 - T^{1, 2}_{x,y}\stretch_1]\frac{\partial\kernel}{\partial\stretch_2}\\
+ [ S^{2, 0}_{x,y}\stretch_0 - T^{2, 0}_{x,y}\stretch_2]\frac{\partial\kernel}{\partial\stretch_0} +  [ S^{2, 0}_{x,y}\stretch_2 - T^{2, 0}_{x,y}\stretch_0]\frac{\partial\kernel}{\partial\stretch_2}.
\end{split}
\end{equation}
Thus we can separate $\tilde{\hessian}^{*}$ into the sum of three parts as shown in \autoref{eqn:three_term_summation_separation}, where each part is defined as
\begin{equation}
\begin{split}
\tilde{\hessian}^{*_{0, 1}}_{x,y} &= [ S^{0,1}_{x,y}\stretch_0 - T^{0, 1}_{x,y}\stretch_1]\frac{\partial\kernel}{\partial\stretch_0} +  [ S^{0,1}_{x,y}\stretch_1 - T^{0, 1}_{x,y}\stretch_0]\frac{\partial\kernel}{\partial\stretch_1},\\
\tilde{\hessian}^{*_{1, 2}}_{x,y} &= [ S^{1, 2}_{x,y}\stretch_1 - T^{1, 2}_{x,y}\stretch_2]\frac{\partial\kernel}{\partial\stretch_1} + [ S^{1,2}_{x,y}\stretch_2 - T^{1, 2}_{x,y}\stretch_1]\frac{\partial\kernel}{\partial\stretch_2},\\
\tilde{\hessian}^{*_{2, 0}}_{x,y} &= [ S^{2, 0}_{x,y}\stretch_0 - T^{2, 0}_{x,y}\stretch_2]\frac{\partial\kernel}{\partial\stretch_0} +  [ S^{2, 0}_{x,y}\stretch_2 - T^{2, 0}_{x,y}\stretch_0]\frac{\partial\kernel}{\partial\stretch_2}.
\end{split}
\end{equation}
If we take a look at one of them, like $\tilde{\hessian}^{*_{0, 1}}_{x,y}$, we have
\begin{equation}
\begin{split}
&\tilde{\hessian}^{*_{0, 1}}_{x,y} = [ S^{0,1}_{x,y}\stretch_0 - T^{0, 1}_{x,y}\stretch_1]\frac{\partial\kernel}{\partial\stretch_0} +  [ S^{0,1}_{x,y}\stretch_1 - T^{0, 1}_{x,y}\stretch_0]\frac{\partial\kernel}{\partial\stretch_1}\\
&= (\omega^{u0}_{x}\omega^{u0}_{y} + \omega^{v0}_{x}\omega^{v0}_{y})\stretch_0\frac{\partial\kernel}{\partial\stretch_0} - (\omega^{u0}_{y}\omega^{v0}_{x} + \omega^{u0}_{x}\omega^{v0}_{y})\stretch_1\frac{\partial\kernel}{\partial\stretch_0}\\
&+ (\omega^{u0}_{x}\omega^{u0}_{y} + \omega^{v0}_{x}\omega^{v0}_{y})\stretch_1\frac{\partial\kernel}{\partial\stretch_1} - (\omega^{u0}_{y}\omega^{v0}_{x} + \omega^{u0}_{x}\omega^{v0}_{y})\stretch_0\frac{\partial\kernel}{\partial\stretch_1}\\
&= \omega^{u0}_{x}(\stretch_0\frac{\partial\kernel}{\partial\stretch_0} + \stretch_1\frac{\partial\kernel}{\partial\stretch_1})\omega^{u0}_{y} - \omega^{u0}_{x}(\stretch_0\frac{\partial\kernel}{\partial\stretch_1} + \stretch_1\frac{\partial\kernel}{\partial\stretch_0})\omega^{v0}_{y}\\
&- \omega^{u0}_{y}(\stretch_1\frac{\partial\kernel}{\partial\stretch_0} + \stretch_0\frac{\partial\kernel}{\partial\stretch_1})\omega^{v0}_{x} + \omega^{v0}_{x}(\stretch_1\frac{\partial\kernel}{\partial\stretch_1} + \stretch_0\frac{\partial\kernel}{\partial\stretch_0})\omega^{v0}_{y}.
\end{split}
\end{equation}
Thus we know $\tilde{\hessian}^{*_{0, 1}}$ has the decomposition as shown in \autoref{eqn:temp_decomposition}. Similar derivations also apply to $\tilde{\hessian}^{*_{1, 2}}$ and $\tilde{\hessian}^{*_{2, 0}}$.

\end{document}